\documentclass{llncs}

\usepackage{graphicx}
\usepackage{amssymb}
\usepackage{amsmath}
\usepackage{url}

\begin{document}
\title{Where is my Device?}
\subtitle{Detecting the Smart Device's Wearing Location\\ in the Context of Active Safety for Vulnerable Road Users}
\author{Maarten Bieshaar}
\institute{Intelligent Embedded Systems Lab, University of Kassel, Germany\\
\email{mbieshaar@uni-kassel.de}}

\maketitle

\begin{abstract}
This article describes an approach to detect the wearing location of smart devices worn by pedestrians and cyclists.
The detection, which is based solely on the sensors of the smart devices, is important context-information which 
can be used to parametrize subsequent algorithms, e.g. for dead reckoning or intention detection to 
improve the safety of vulnerable road users. 
The wearing location recognition can in terms of \textit{Organic Computing (OC)} be seen as a step towards 
self-awareness and self-adaptation. 
For the wearing location detection a two-stage process is presented. It is subdivided into 
moving detection followed by the wearing location classification. Finally, the approach is evaluated 
on a real world dataset consisting of pedestrians and cyclists. 

\end{abstract}


\section{Introduction}
\label{mb:sec:introduction}

For future traffic scenarios, we envision autonomously driving cars, trucks, and other road users
equipped with sensors, electronic maps, interconnected via a Car2X-network, allowing 
for cooperation on different levels, such as situation prediction or intention detection~\cite{mbieshaarBRZ+17}.
Each road user determines and continuously maintains a local model of the surrounding traffic situation which is enhanced by  information and models originating from other road users, intelligent infrastructure~\cite{mbieshaarGSM+12}
and vulnerable road users (VRU, i.e. cyclists and pedestrians) themselves if equipped with mobile phones and wearables, such as smartphones and smartwatches.
VRUs will play an important role in future traffic. 
To avoid accidents and achieve a highly efficient traffic flow, it is important to detect VRUs and to predict their intentions.
The collective intelligence of all road users, infrastructure, and VRU are used to reach for a cooperative 
VRU intention detection approach~\cite{mbieshaarBie16} with the overall goal of increased safety, 
especially the safety of VRU.
The vision follows the principles of \textit{Organic Computing (OC)}~\cite{mbieshaarMSU11} systems and research, in which 
a self-organized decentralized system consisting of autonomous entities which all perceive, act locally, cooperate, and solve complex tasks. 

In our envisioned scenario the VRU itself becomes a valuable source of information when it comes to 
detecting its presence, predicting its intentions, and emitting a warning to the driver of an approaching 
car. Smart devices are equipped with various sensors such as accelerometers, gyroscopes, compasses, and even GPS receivers. Although these body-worn sensors and smart devices are 
capable of detecting the coarse movements, their directions, or even the persuaded activity (e.g. walking or cycling), they are limited concerning their position and velocity measurement accuracy. 
Especially in urban areas the GPS signal is not always present, nor is the positional accuracy vast enough. 
Hence, in the absence of the GPS signal other localization techniques are required. Besides others, such as WiFi fingerprinting~\cite{mbieshaarVD16}, 
the integration of additional smart device based aided inertial system or dead reckoning, e.g. tracking of VRU position by integration of velocity estimates, can be used to determine the position of the VRU~\cite{mbieshaarEng13}. 

One central aspect influencing the accuracy of smart devices activity detections as well as 
the positional accuracy of the aided inertial system is the device wearing location.
In~\cite{mbieshaarVAS11}, it was shown that the performance of a pedometer (the basis for pedestrian 
dead reckoning) varies tremendously, e.g. when placed in the trouser pocket or at chest. Furthermore, 
many algorithms especially for pedestrian orientation estimation make explicit use of the wearing location, e.g. 
leg and arm sway forth and back~\cite{mbieshaarBS08}.

In this article, we adopt the following distinction between wearing position and location~\cite{mbieshaarOgr09}.
A position refers to place given in absolute coordinates within a particular reference frame, whereas 
location refers to some abstract spatial information, e.g. wrist or chest. Our approach allows to detect 
predefined wearing locations.

In terms of organic computing, the system requires self-reflection capabilities in order to estimate how 
certain it can be about its predictions, i.e. detected intentions or predicted position, 
or when to adapt and to use different filter algorithms, e.g. for dead reckoning. 
In this article an approach to smart devices wearing location detection is presented, 
aiming to bring self-reflection capabilities and context awareness to smart devices in the domain of 
VRU intention detection.

\subsection{Related Work}
\label{mb:sec:related_work}

Kunze~et~al.~\cite{mbieshaarKLJ+05} were about the first to present an approach to detect the wearing location of smart devices.
Their approach first detects the pedestrian walking pattern and then subsequently estimates the wearing location.
Vahdatpour~et~al.~\cite{mbieshaarVAS11} studied on the on-body location to determine optimal sensor placement for activity recognition. 
As with Kunze~et~al. their wearing location detection approach first detects pedestrian walking motion and then in the subsequent step classifies the device location using a support-vector machine. However, their results suggest that detection of six main regions, i.e. forearm, head, shin, thigh, upper arm and waist, is adequate for most activity recognition applications.
In~\cite{mbieshaarSS16}, Sztyler and Stuckenschmidt investigated on wearing location-aware activity recognition. 
Their approach consist of multiple stages: first a classification between static and dynamic activities is performed, 
then the smart device location is determined using a random forest and finally a location-aware classification is performed.
They showed that location-awareness can help to increase activity recognition performance.
In~\cite{mbieshaarPPC+12}, Park~et~al. presented an approach to online wearing location assessment and velocity 
estimation based on support-vector regression. Their system showed promising results but they evaluated it only for pedestrians.
In~\cite{mbieshaarEng13}, the authors propose an approach for improved smart device based positioning for active VRU protection based 
on Car2Pedestrian communication. The authors pointed out, that one central factor limiting the applicability of their approach is the system's dependence on the wearing location of the smart device, i.e. they assumed for their system that the smartphone is located in the front trouser pocket. They were not able to directly transfer their trained classifiers to other wearing locations. This stresses 
the importance of being able to detect the smart device wearing location.
In~\cite{mbieshaarCha14}, Chang considered different wearing location when building a system to accurately localise cyclists using GPS and dead reckoning. His approach first detects the smart device wearing locations and then uses this context information to estimate the 
pedal frequency in order to get a coarse velocity estimate.

\subsection{Main Contributions and Outline}
\label{mb:sec:outline}

The main contribution of this article is an approach to detect the smart device's wearing location in the context of active safety for vulnerable road users is presented. 
For detection machine learning and human activity recognition techniques are applied.
These are evaluated not only for pedestrians but also for cyclists.

The remainder of this article is structured as follows: Section~\ref{mb:sec:method} 
describes the methods used for detecting the smart device's wearing location. In the following Section~\ref{mb:sec:evaluation}, 
the evaluation of the approach, including the evaluation metric and the conducted experiments using pedestrians and cyclists 
is described. Finally, in Section~\ref{mb:sec:conclusion} the main conclusion and open research questions for future work are reviewed.

\section{Method}
\label{mb:sec:method}
\begin{figure} [h]
	\centering
	\includegraphics[width=\textwidth, clip, trim=10 40 10 110]{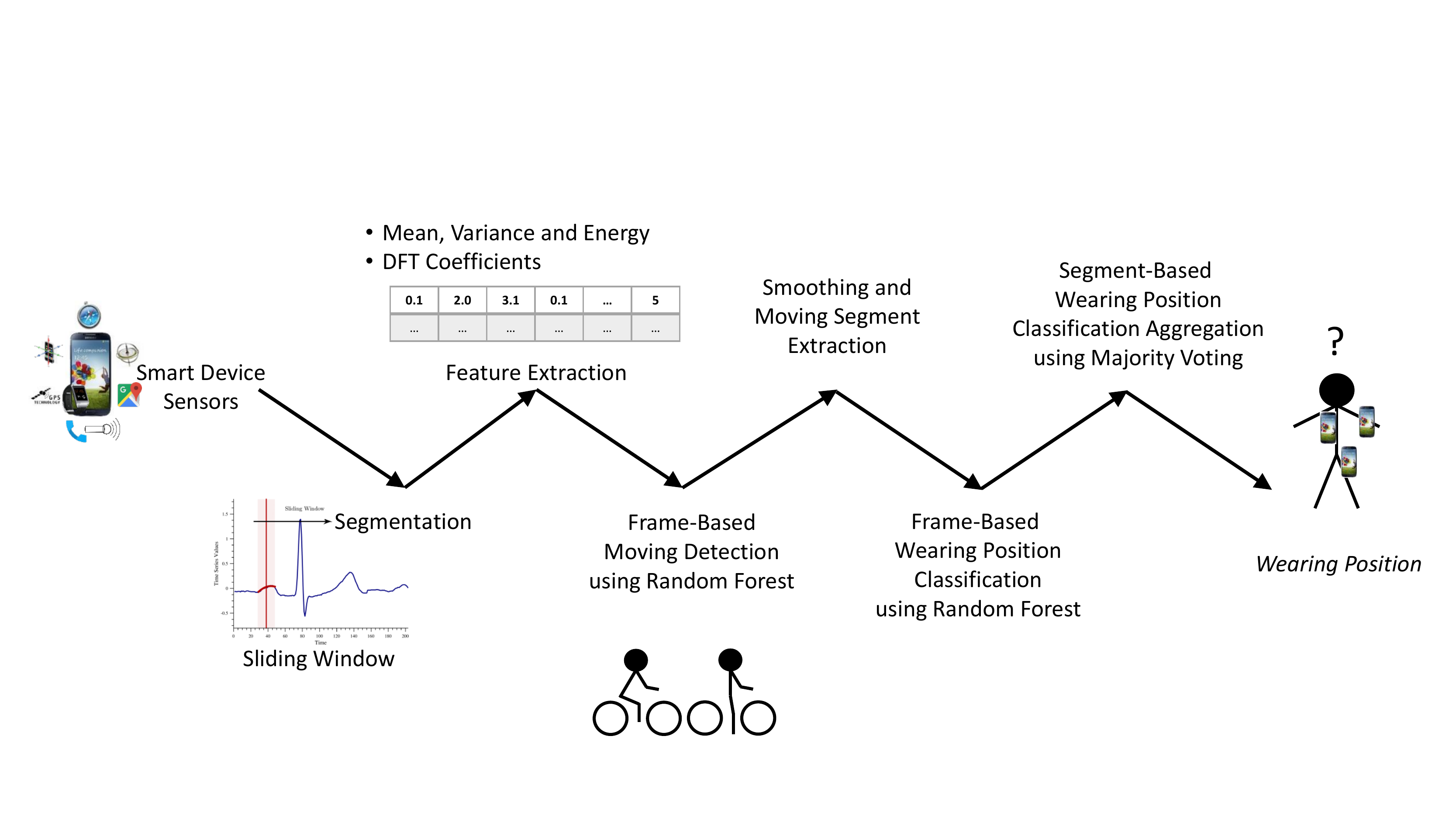}
	\caption{Process of wearing location recognition based on smart device activity data.}
	\label{fig:recognition_pipeline}
\end{figure}

In this Section the methodology for smart device wearing location detection is presented.
The approach presented in this article does only consider pedestrians and cyclists. Neither the classification between the different VRU types nor the influence on classifying the 
VRU class wrong is in the scope of the article and is left open for future research.
Therefore, in the following we assume that the class of the VRU is given. 
The wearing location detection approach consists of a two-stage process, 
which is inspired by the work of Kunze~et~al.~\cite{mbieshaarKLJ+05}. 
In the first stage a two-class classification between standing and moving is performed. 
The goal of this classification is to filter out non-informative patterns. If the pedestrian or cyclist is standing and 
not moving it is hardly possible to identify the current wearing location. This classification is referred to as moving 
detection. Then in the subsequent stage the wearing location is determined. It is triggered by the moving detection. 
Both, the process of moving detection and the wearing location recognition follow the steps of the human activity recognition pipeline~\cite{mbieshaarBBB14}. The complete process is visualized in Fig.~\ref{fig:recognition_pipeline}.

\subsection{Preprocessing and Feature Extraction} 
\label{mb:subsec:preprocessing_feature_extraction}

The system uses as input accelerometer and gyroscope data which is resampled at  $100$~Hz. Subsequently the three accelerometer 
and gyroscope components ($x, y$ and $z$, respectively) are transformed using the estimated gravity vector (e.g. obtained by low-pass 
filtering of accelerometer measurements), such that in the resulting coordinate frame the components are levelled with the local earth ground plane, i.e. the $z$-axis is pointing towards the center of gravity. 
This coordinate frame is referred to as local frame. In absence of any compass data or other reference data, the exact transformation from this local coordinate frame to the local coordinate frame of the VRU is unknown, i.e. we do not know how the device is oriented with respect to the VRU. In order to compensate for this, the magnitudes of the projected accelerometer and gyroscope values in the local horizontal $x-y$ plane are computed because the magnitude is invariant against rotation around the local vertical $z$-axis. 
Additionally, the projection of the input data on the gravity axis is considered. This reduces the input dimension from three values per sensor ($x$, $y$, and $z$) to two values per sensor ($x-y$ plane and gravity axis) and time unit. 

A sliding window segmentation is performed on these transformed values. It is used to compute mean as input feature of each signal. Moreover, similar to~\cite{mbieshaarPPC+12} magnitudes of the discrete Fourier transform (DFT) coefficients are also considered as input features. The DFT coefficients are computed over a window of $2.56$~s. The window is long enough to include a full movement cycle, e.g. gait- as well as pedalling cycle. Under consideration of the sample frequency ($100$~Hz) this corresponds to a distance between spectral components of $~0.39$~Hz. Human movement is captured well by lower frequencies and, therefore, only the first $30$ spectral components are kept (up to $11.7$~Hz). Moving patterns, such as walking or cycling, are mainly represented in the spectral intensity spectrum, 
while the shape of the Fourier spectrum can be used for detection of the wearing location.

Moreover, before computing the DFT coefficients the DC component, i.e. mean value of the current window is subtracted from each sample. In order to make the spectral magnitude independent of the overall energy in the considered window, the coefficients' magnitudes are normalized with respect to the energy~\cite{mbieshaarPPC+12}. Additionally, the approximation errors of the signals' Fourier approximation are considered as input features. The DFT approximation is computed using a fast and efficient incremental sliding window algorithm~\cite{mbieshaarGGS13}.
This results in $32$ features extracted from the DFT approximation, i.e. $30$ coefficients' magnitudes, bias, and approximation error. Considering the transformed accelerometer and gyroscope measurments, and the additional mean feature for each sensor, this results in total in $132$ ($4 \times 33$) features.

\subsection{Moving Detection} 
\label{mb:subsec:standing_moving_detection}
A robust moving detection is essential for wearing location classification. The moving detection is realized by means of a frame-based classifier being trained to distinguish between moving and standing frames. 
The classifier is triggered twice a second, i.e. at $2$~Hz.
A random forest~\cite{mbieshaarBre01} classifier is trained on the previously extracted features.
The random forest includes by construction a simple yet effective feature
selection mechanism making it able to cope with the high dimensional feature space. The hyper-parameter of the random forest are tuned using a coarse-to-fine grid search and five-fold cross-validation over the test subjects. 

As with any classification task, the trade-off between precision and recall has to be tackled. Here, we favour to avoid false detection such that a potential negative influence of wrong moving detection on the subsequent carrying classification stage is minimized. In order to reduce 
the false positive rate, a subsequent smoothing stage is added to the moving detection. It consists of a sliding window of length $2.5$~s aggregating 
the last five predictions. A positive moving classification is only emitted 
if $80\%$ of the aggregated predictions are moving. The frame-based moving predictions are then further aggregated to segments of variable length, allowing for segment-based wearing location classification.

\subsection{Wearing Location Classification}
\label{mb:subsec:wearing_location_classification}

The wearing location classification is performed only for 
frames which are classified by the preceding stage as moving.
The frame-based wearing location classification is again realized 
using a random forest classifier using the features as described in Section~\ref{mb:subsec:preprocessing_feature_extraction}. The classification 
is performed at a frequency of $2$~Hz. The frame-based classification is further improved by an ensemble approach aggregating all classification within a moving segment. The aggregation is realized using a majority voting schema.
 Additionally, a minimal segment length of $2.5$~s is enforced. 
 This is based on the observation that errors in the moving classification 
 are rather short. Hence, enforcing a minimal length avoids 
 errors from the moving detection to propagate into the wearing location 
 classification. The parameters of the classifier are optimized using a coarse-to-fine grid search and five-fold cross-validation over the test subjects.

\section{Evaluation}
\label{mb:sec:evaluation}

In this Section the approach to smart devices wearing location is evaluated. The evaluation of the classifier's performance 
is based on the confusion matrix, the $F1$ score (the harmonic mean between precision and recall) and 
related scores for human activity recognition~\cite{mbieshaarMWW+06}.

\subsection{Data Acquisition}
\label{mb:sec:data_acquisition}

The developed algorithms are evaluated in experiments conducted with 49 female and male test subjects, all being in the age 
between $18 - 54$. The test subjects were instructed to move between certain points at a research intersection~\cite{mbieshaarGSM+12} 
with real traffic. Furthermore, they were asked to follow the traffic rules.
Activity data from each of the test subjects was recorded for walking on foot, i.e. being a pedestrian, and using a bicycle. 
The test subjects were equipped with four smartphones, worn at different body locations.
The smart devices used for evaluation are Samsung Galaxy S6 smartphones.
For both, the pedestrian and the cyclist four smart device wearing locations are investigated.
Both, the pedestrian as well as the cyclist were equipped with smartphones in their front trouser pocket and 
front pocket of their jacket at the height of their chest. Moreover, the test subjects were also equipped with 
a rucksack which also contained a smartphone. For those test subject which did not posses a front pocket at the chest, the smartphone was mounted within a pocket located at chest belt of the rucksack. The pedestrians carried the fourth smartphone in the back trouser pocket, whereas the cyclists carried their fourth smartphone in a pocket mounted at the rack of the bicycle. The smartdevices were all placed in pockets in a predefined orientation, e.g. for front trouser pocket upright position and display facing outwards. 
This is not restricting the general applicability of the presented approach, but rather makes the experimental setup more comprehensible and reproducible. Furthermore, different bicycles were used during the experiments, ranging from mountain bikes over city bikes to racing bikes.

For both cases, pedestrian as well as cyclist, distinct models, i.e. moving detection and wearing location classification, are trained and optimized. 
The evaluation is also performed separately for both cases using a 
nested cross-validation over the test subjects. As described in Section~\ref{mb:sec:method} the inner five-fold cross-validation is used for model validation, whereas the outer five-fold is used for testing.
In total the evaluation is based on $11$ pedestrians and $52$ cyclists.

\subsection{Wearing Location Classification}
In this section the results of the smart devices wearing location classification are presented. Therefore, an overall evaluation 
of the classifier's performance over all test subjects is performed. Pedestrians and cyclists are considered individually.
The classification statistics, such as precision, recall and F1-Score are 
calculate for the multi-class classification problem for each label separately and then averaged weighted by the support of each class.

\begin{figure}[!htb]
	\begin{minipage}{0.3\textwidth}
		\vspace{-2cm}
		\begin{tabular}{ | l | l|}
			\hline
			Score \hspace*{0.5cm} & Evaluation \hspace*{0.4cm} \\ \hline \hline
			Accuracy & \hspace{0.1cm} 0.795 \\ \hline
			Precision & \hspace{0.1cm} 0.856  \\ \hline
			Recall & \hspace{0.1cm} 0.795 \\ \hline
			F1-Score & \hspace{0.1cm} 0.789 \\
			\hline
		\end{tabular}
	\end{minipage}%
	\begin{minipage}{0.7\textwidth}
		\flushright
		\includegraphics[width=\linewidth, clip, trim=30 10 10 25 ]{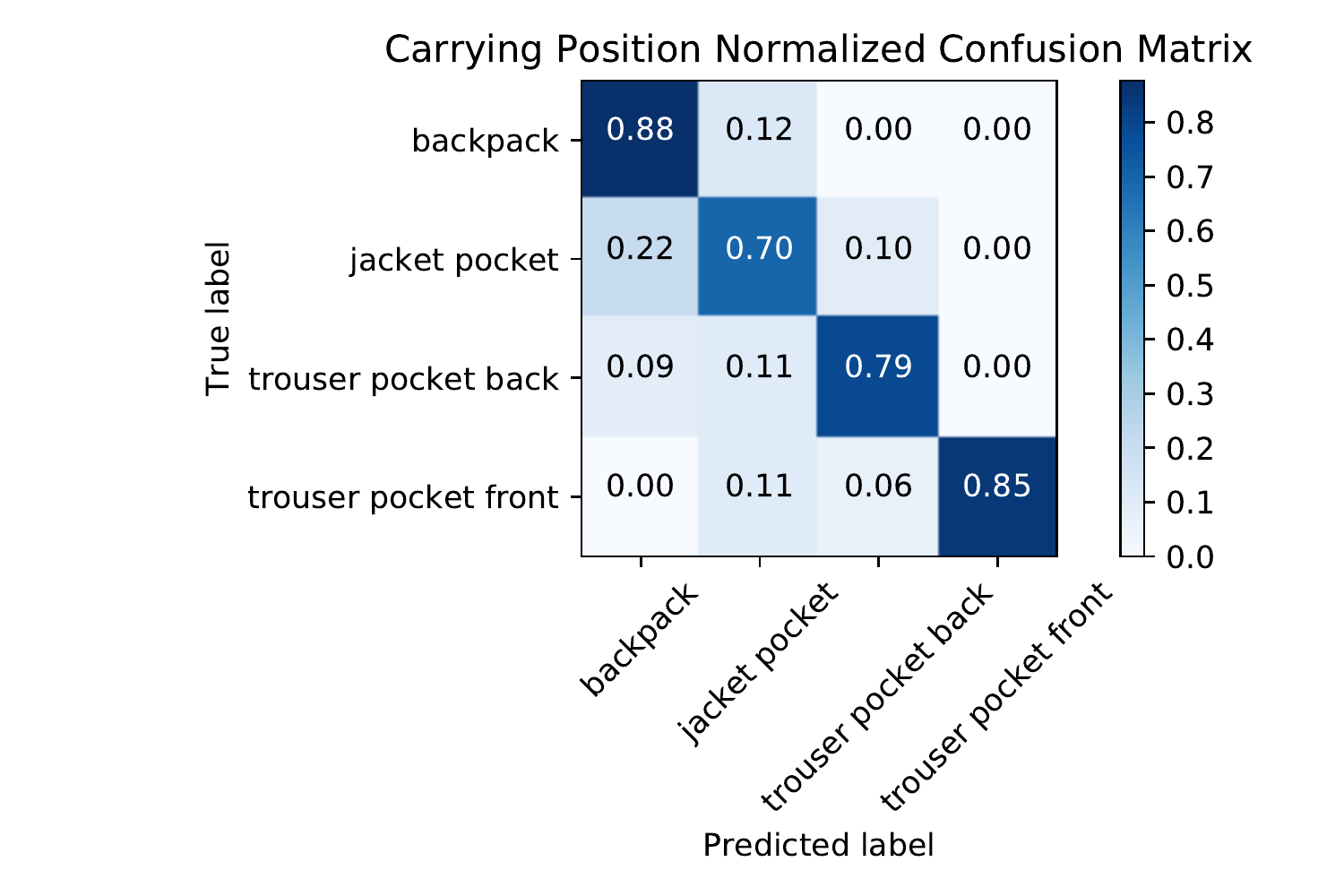}
	\end{minipage}
	\caption{Results pedestrian segment-based wearing location classification.}
	\label{fig:pedestrian_results}
\end{figure}

In case of wearing location classification for devices worn by pedestrians 
on average an accuracy of $0.795$ over all segments is achieved. 
As it can be seen in Fig.~\ref{fig:pedestrian_results}, the precision 
is even higher, showing that there a few false positive predictions.

Inspecting the classification results in the confusion matrix, as depicted in 
Fig.~\ref{fig:pedestrian_results}, more closely, we see that the trouser pocket front wearing location classification score, i.e. precision, is high. Surprisingly, there are only little confusions between 
the trouser pocket back and front position. Although, they are 
concerning the body location close together, the activity pattern 
seems to be different, such that a safe classification between both is 
possible. But instead a classification between the jacket pocket and 
the trouser pocket back is harder to realize, and more error 
prone. The backpack wearing location is often confused with the 
jacket pocket which is most likely due to a very similar location 
and observed movement pattern with respect to the human body.

\begin{figure}[!htb]
	\begin{minipage}{0.3\textwidth}
		\vspace{-1.5cm}
			\begin{tabular}{ | l | l|}
				\hline
				Score \hspace*{0.5cm} & Evaluation \hspace*{0.4cm} \\ \hline \hline
				Accuracy & \hspace{0.1cm} 0.905 \\ \hline
				Precision & \hspace{0.1cm} 0.910  \\ \hline
				Recall & \hspace{0.1cm} 0.905 \\ \hline
				F1-Score & \hspace{0.1cm} 0.905 \\
				\hline
			\end{tabular}
	\end{minipage}%
	\begin{minipage}{0.7\textwidth}
		\flushright
		\includegraphics[width=0.95\linewidth, clip, trim=30 10 12 25 ]{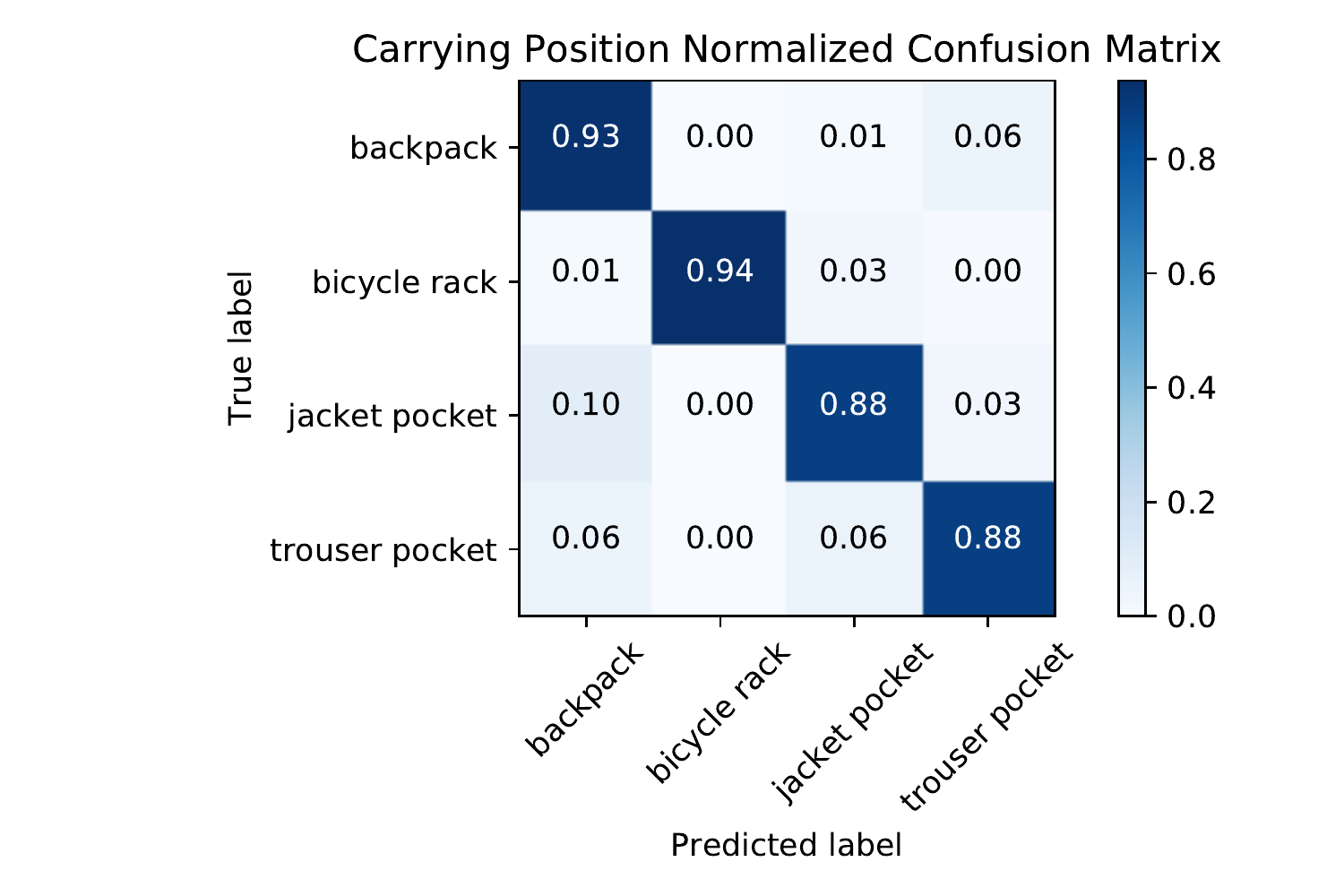}
	\end{minipage}
	\caption{Results cyclist segment-based wearing location classification.}
	\label{fig:cyclsit_results}
\end{figure}

The results for cyclists' the smart device wearing location classification 
are presented in Fig.~\ref{fig:cyclsit_results}. Here an average accuracy 
score of $0.905$ is reached. As it can bee seen in the confusion table 
the separate wearing locations can be well distinguished from each other. As with the pedestrians the backpack is sometimes confused with the jacket pocket. In a few cases also the trouser pocket (front) is confused with the jacket pocket. This is most likely caused by some test subjects 
wearing their smart phone in the lower jacket pocket, which is concerning the human body while cycling close to the trouser pocket wearing location.

Although the technique applied here is similar to the method presented by Kunze~et~al.~\cite{mbieshaarKLJ+05}, we were not able to reach a similar performance concerning pedestrian smart device wearing location classification. This is most likely due to different and harder distinguishable wearing locations. It can be seen in the confusion tables that most classification errors are due to similar wearing locations. Moreover, the dataset used for evaluation 
is different. It contains more pedestrians but the walking sequences 
are shorter. Hence, the total amount of walking time is smaller, increasing the influence of single errors. The results for the cyclists smart device wearing location are much better showing the functionality and applicability of the approach.

\section{Conclusion and Outlook}
\label{mb:sec:conclusion}

In our preliminary work on creating an aided inertial navigation system for accurate VRU localization, we detected a strong 
relationship between on the one hand side the wearing location and on the other hand side the noise level of the acceleration and gyroscope sensor values and the GPS accuracy. In this work, we presented an approach to detect the smart devices wearing location 
in order to improve context-awareness and parametrise subsequent algorithms, e.g used for dead reckoning. 

We were able to show, that the approach can classify the correct wearing location with an accuracy of approx. $90$\% for cyclists and $80$\% for pedestrians. Although this is still far from perfect classification, it may concerning the use case of helping to parametrise subsequent systems, e.g. an inertial navigation, already be fair enough. Nevertheless, further improvements could potentially be achieved using other classifiers such as neural networks, more sophisticated ensemble approaches and more training data.

Additionally, in future work an interesting point is how long the classifier needs until it can deliver classification results with a desired certainty.  

Our future work will focus on using the information obtained from the wearing location estimation procedure in order to develop an  
aided inertial navigation system based on the smart devices sensors and GPS. The positional accuracy of this system shall 
allow to perform VRU intention detection, solely based on smart device data. Moreover, we also aim to investigate 
on the relationship between smart device movement primitive detection~\cite{mbieshaarBZD+17} and the smart device wearing location.

\section*{\large Acknowledgment}

This work results from the project DeCoInt$^2$, supported by the German Research Foundation (DFG) within the priority program SPP 1835: "Kooperativ interagierende Automobile", grant number SI 674/11-1.

\bibliographystyle{splncs}
\bibliography{mb_literature}

\end{document}